\documentclass[12pt]{iopart}

\usepackage{natbib}  
\usepackage{graphicx}
\usepackage{hyperref}
\begin{document}

\title[Kabath et al. 2019]{Ond\v{r}ejov echelle spectrograph, ground based support facility for exoplanet missions \footnote{This article is based on the data collected with Perek 2-m telescope.}}

%\author{Content \& Services Team}

 \author{P. Kab\'{a}th$^1$, M. Skarka$^1$, S. Sabotta$^2$, E. Guenther$^2$, D. Jones$^{4,5}$, T. Klocov\'{a}$^1$, J. \v{S}ubjak$^1$, Ji\v{r}\'{i} \v{Z}\'{a}k$^{3,1}$, M. \v{S}pokov\'{a}$^{1,3}$, M. Bla\v{z}ek$^{1,3}$, J. Dvo\v{r}\'{a}kov\'{a}$^1$, D. Dupkala$^1$, J. Fuchs$^1$, A. Hatzes$^2$, E. Kortusov\'{a}$^1$, R.  Novotn\'{y}$^1$, E. Pl\'{a}valov\'{a}$^6$, L.  \v{R}ezba$^1$, J.  Sloup$^1$, P. \v{S}koda, M. \v{S}lechta$^1$
         }  
    
\address{$^1$ Astronomical Institue, Czech Academy of Sciences, 
              Fri\v{c}ova 298, 25165, Ond\v{r}ejov, Czech Republic\\}
\address{$^2$ Th\"{u}ringer Landessternwarte Tautenburg, Sternwarte 5, 07778 Tautenburg, Germany \\}
\address{$^3$ Department of Theoretical Physics and Astrophysics, Masaryk Univesity, Kotl\'{a}\v{r}sk\'{a} 2, 60200 Brno, Czech Republic \\}
\address{$^4$ Instituto de Astrof\'isica de Canarias, E-38205 La Laguna, Tenerife, Spain Departamento de Astrof\'isica, Universidad de La Laguna, E-38206 La Laguna, Tenerife, Spain}
\address{$^5$ Universidad de La Laguna, E-38206 La Laguna, Tenerife, Spain}
\address{$^6$Mathematical Institute SAS, Štefánikova 49, 814 73 Bratislava, Slovakia}
\ead{petr.kabath@asu.cas.cz}     

\vspace{10pt}
\begin{indented}
\item[]Sepotember 2019
\end{indented}

\begin{abstract}

Fulfilling the goals of space-based exoplanetary transit surveys, like Kepler and TESS, is impossible without ground-based spectroscopic follow-up.  In particular, the first-step vetting of candidates could easily necessitate several hundreds of hours of telescope time -- an area where 2-m class telescopes can play a crucial role.  Here, we describe the results from the science verification of the Ond\v{r}ejov Echelle Spectrograph (OES) installed on the 2-m Perek telescope.  We discuss the performance of the instrument as well as its suitability for the study of exoplanetary candidates from space-based transit surveys.  In spite of being located at an average European observing site, and originally being conceived for the study of variable stars, OES can prove to be an important instrument for the exoplanetary community in the TESS and PLATO era -- reaching accuracies of a few tens of m/s with reasonable sampling and signal-to-noise for sources down to V$\sim$13.  The stability of OES is demonstrated via long-term monitoring of the standard star HD~109358, while its validity for exoplanetary candidate verification is shown using three K2 candidates EPIC~210925707, EPIC~206135267 and EPIC~211993818, to reveal that they are false positive detections.

\end{abstract}

\vspace{2pc}

%\submitto{\PASP}

%\maketitle

\section{Introduction}

The {\it Kepler} space mission discovered about $2500$ confirmed exoplanets. The {\it Kepler}/K2 mission uses the method of transit detection, i.e. the measured decrease in the stellar flux when an exoplanet moves in front of its host star \citep{borucki16}. However, there are still several thousands of candidates, needing confirmation by other methods. Even for validated planets -- ones for which we are confident of their planetary nature -- spectroscopic measurements are still needed to determine the planet mass. For non-validated planets, false positive detections have various causes such as background binary stars, incorrect spectral type determination and grazing eclipses, with the rate of such false positive detections being discussed already in time of the first ground based exoplanet searches  \citep{brown03}. Therefore, systematic ground-based follow-up of all candidates is crucial in order to separate candidates from true exoplanetary systems as well as to derive the parameters of the systems \citep{cabrera17, shporer17}.
% as first shown by the CoRoT space mission \citep{auvergne09}. The need for thorough independent ground based follow-up observations is nicely illustrated on three K2 planets which turned out to be background eclipsing binaries \citep{cabrera17} or Three Statistically Validated K2 Transiting Warm Jupiter Exoplanets Confirmed as Low-mass Stars \citep{shporer17}.

There are several important steps in the process of confirming transit candidates. First, photometric follow-up is needed to confirm that the transit is on-target and not due to a contaminating star in the aperture of the space instrument \citep{deeg09}. Second, the spectral type of the host star gives the first estimate of the stellar radius and thus planet radius which will eliminate stellar binary companions \citep{guenther12, sebastian12, gazzano13, damiani16, guenther10, ammlervoneif15}. Third, a few RV measurements with low precision (100--1000 m/s) can then be used exclude the eclipsing binary scenario. Finally, those candidates which pass the initial screening can be followed up by instruments capable of precise RV measurements \citep{bouchy09, loeillet08, gandolfi10, leger09}, such as HARPS, UVES or ESPRESSO, for the determination of the planet mass. Given the intrinsic brightness of the candidates discovered by space-based missions, all of the tasks preceeding the final high-precision RV measurements with large facilities can be carried out with a 2-m telescope equipped with appropriate instrumentation. Furthermore, Jupiter-sized planets and brown dwarfs showing RV semi-amplitudes of hundreds of m/s can be even characterised with such telescopes \citep{endl02, dollinger07}, eliminating the need for large amounts of observing time with more over-subscribed facilities. In short, an efficient follow-up effort can and should employ 2-m class telescopes in order to ease the burden on large facilities where telescope resources are scarce.

The initial false positive rate for CoRoT was about  83\,$\%$, with the remaining 17\,$\%$ requiring follow-up observations, from which 12\,$\%$ were confirmed as planets  \citep{almenara09}. The false positive rate for {\it Kepler}/K2 was reported to be around $10$\,$\%$ for all candidates and around $20$\,$\%$ for hot-Jupiters \citep{santerne12, fressin13}. These false positive rates were confirmed by numerous other studies  \citep{lillo-box14, dressing14, sliski14, colon12}. A detailed study by  \citet{santerne13} shows that, due to the large TESS PSF of 21 arcsec, the removal of false positives by using ground-based facilities is even more important than for other missions. Since the PSF of TESS is substantially larger than the Kepler PSF and the duration of a typical light curve is much shorter than Kepler's four years, the centroid method (Section 3.2) is far less efficient. Using the calculations by  \citet{santerne13} we expect a statistical field contamination between 0.2 and 20 stars within the PSF.

Ground based follow-up will also be extremely important for the PLATO space mission, expected to be launched in $2026$, which will require more than $50$ nights of observing time per year for six years on 2-m class telescopes  to characterise most of the detected candidates \citep{red}. Therefore, the importance of 2-m class facilities with a lot of available observing time with state-of-the-art instrumentation is of growing importance. 

In this article, we describe first results of scientific verification process of the Ond\v{r}ejov \'{E}chelle Spectrograph (OES), which is installed on the 2-m Perek telescope located in the Czech Republic. Furthermore, we also present the first results of K2 candidates follow-up using the instrument. In Section \ref{sec:Ondrejov}, the observing site and the telescope will be described. In Section \ref{sec:OES} we present details of the OES instrument, while in Section \ref{sec:RV} we discuss the instrumental stability and the RV precision obtained during science verification. We conclude with a discussion of the potential for exoplanetary research with the OES in Section \ref{sec:conc}.

\section{Perek Telescope, ground based support of space missions} \label{sec:Ondrejov}

The facility is located 30 kilometers south-east of Prague. The telescope itself is maintained and operated by the Astronomical Institute of Czech Academy of Sciences and it is considered as a national facility, therefore, researchers from all Czech institutions have access to the telescope. Furthermore, under certain conditions, international projects are also accepted. The Perek telescope has a primary mirror with a diameter of 2 meters and is equipped with two instruments, a single-order spectrograph and an \'{e}chelle spectrograph. The telescope was inaugurated in 1967 and it is a twin telescope of the Alfred Jensch telescope located in Thuringia in Germany. 

One of the instruments of Perek telescope is OES -- the subject of our science verification campaign described in the following sections. The instrument was built in Ond\v{r}ejov in $2007$ but it saw little use due to lack of scientific interest of the observatory staff. However, renewed interest in OES instrument appeared in April 2017 after establishing a new exoplanetary research group at the Academy and gaining access to an iodine cell. Currently, OES is primarily used for exoplanetary ground based follow-up projects, such as discrimination of false positives from space missions and for spectral typing of exoplanetary candidates host stars. In addition, we are pursuing a long term monitoring project of exoplanets around A-type stars in collaboration with Tautenburg Observatory. \newline 

\noindent {\bf Observing site quality} \newline

\begin{figure}
   \centering
   \includegraphics[width=10cm]{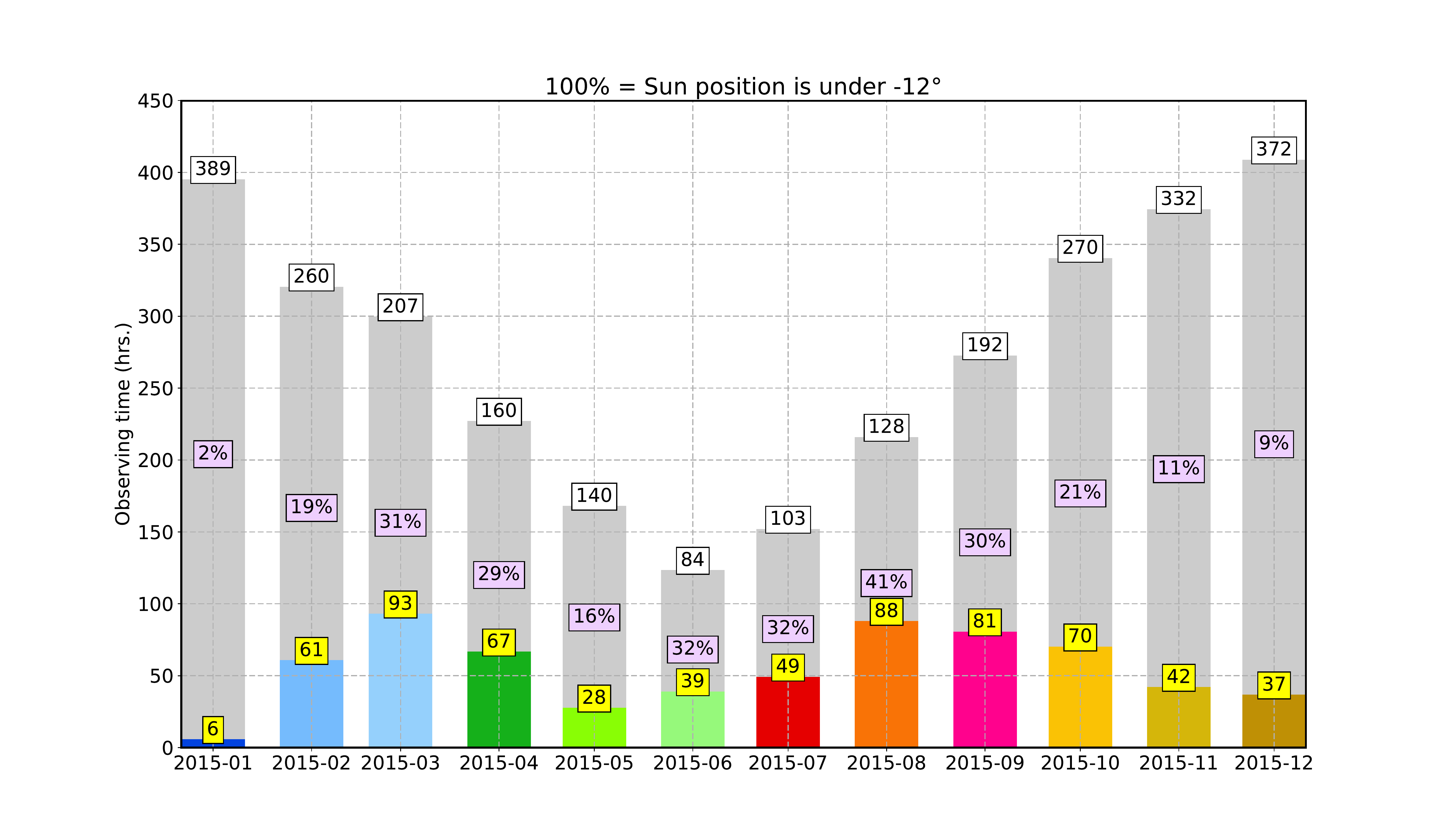}
      \caption{Typical yearly distribution of observing hours over months in the year. The grey column is a total number of observable hours (Sun below $-$12 deg), the coloured number is the actual total of observed hours and the percentage gives the ratio of total available and observed hours.   
              }
         \label{weather}
   \end{figure}

\noindent The observing site has a characteristic central European weather pattern. Fig.\ref{weather} shows the distribution of hours available for observing (Sun below 12 deg) versus actually observed hours over the year 2015. Typically, the winter period from October through February has less favourable conditions for observing characterised with high humidity and low level clouds. This is reflected in the comparison of the percentages of available time that was actually observed in winter (October--February), being only a few percent, and during summer (March--September), being an order of magnitude greater at $30--40$\,$\%$. We are able to obtain good data on average of $25--30$\,$\%$ of all available observing hours during the whole year.

Typical seeing for Ond\v{r}ejov is between 2--3" and under excellent conditions it can be as good as 1.5". However, for our new key program, ground based follow-up of exoplanetary candidates, we are generally not limited by the seeing. The limiting magnitude is currently about 13\,mag in {\it V}. We are able to obtain a spectrum of SNR $7$ for a 12.6 mag V star in $1.5$ hours exposure. 

\section{Ond\v{r}ejov Echelle Spectrograph---OES}\label{sec:OES}

The spectrograph of OES is installed in a temperature stabilised room and is fed by the telescope via the Coude focus. OES is a white-pupil spectrograph. The light path from the telescope to OES has a 5 mirror train bringing the light to a focus on a $0.6$\,mm slit (corresponding to 2") and subsequently to the spectrograph in a separate room. The slit-viewer camera, a TV EEV CCD $65$, is installed in the control room in a closed box along with the Iodine cell. The Iodine cell can be moved in the light beam on demand. Standard calibrations for science frames can be performed using the flat field lamp and ThAr lamps which are mounted beside the optical path where a mirror can be inserted such that the beam from the lamp(s) goes to the spectrograph. (unfortunately, no simultaneous ThAr calibrations are possible). Dome flats can be taken using a white screen installed in the dome.  

   \begin{figure}
   \centering
   \includegraphics[angle=0,  width=8cm, height=5.8cm]{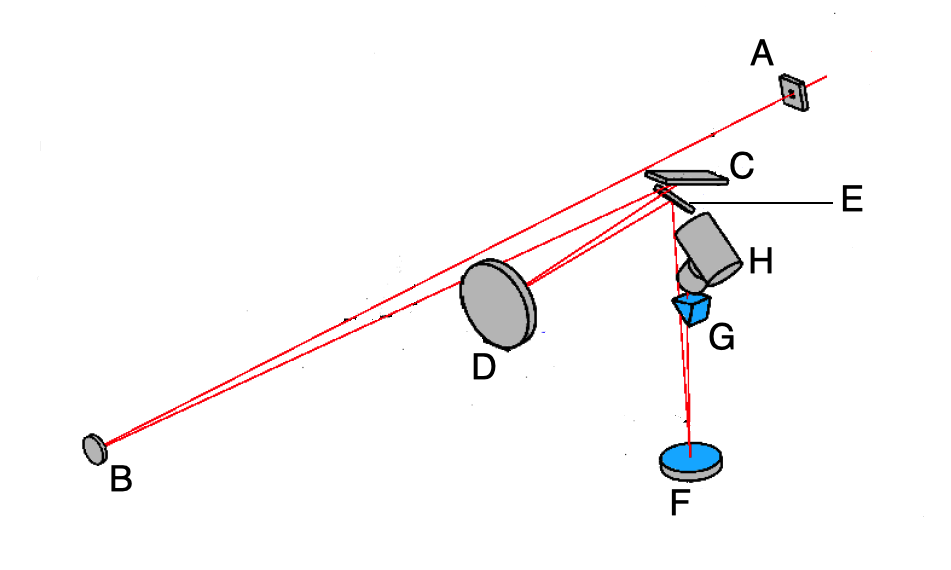}
 %  \includegraphics[angle=0,  width=8cm, height=5.8cm]{IMG_9469.JPG}
  %    \caption{Upper panel: Spectrograph design. Figure from \citet{koubsky04}. Bottom panel: \'Echelle grating, cross-disperser and the detector.
  \caption{OES light comes from the Coud\'{e} room through the slit A to collimator B. From the collimator the light beam travels to an \'{e}chelle grating C and later to a parabolic mirror D and a plane mirror E. Second collimator F is in front of the cross-disperser which is the last element before the CANON lens objective H with a detector. Courtesy of Mirsolav \v{S}lechta.
              }
         \label{spectro}
   \end{figure}

The spectrograph itself is mounted on the stable construction as presented in more detail in \cite{koubsky04}. The incoming light from the telescope travels to a collimator mirror of 147 mm diameter (B in Fig \ref{spectro}). The light is then reflected to an \'{e}chelle grating of $408\times165\times 74$\,mm with $54.5$ g/mm and a blaze angle $\theta = 69^\circ$ (C). The optical mirror (F) reflects the light to a LF5 cross-disperser prism (E) which separates the orders that are then brought to a focus on the detector using a Canon EF 200 f/1.8 objective (I). The detector is an EEV $2048\times2048$ pixel, nitrogen-cooled CCD with a pixel size of $13.5$ $\mu$m, a dynamical range of 65,000 ADUs, a readout noise of about 3.5 $e^-$ RMS and a dark current of 1 $e^-$/p/hr. Parameters of optical elements of the spectrograph can be found in Table\,\ref{opt}. The wavelength coverage of the spectrograph is from 3750 to 9200 \AA, with a resolving power of about $R=50,000$ and spectral sampling 2.4\,\AA/mm. The spectral range is covered by 56 usable orders which overlap in the blue about $40$\,\AA but there are gaps in the wavelength coverage of up to $40$\,\AA in red orders starting at about $4500$ \AA. The guiding of the star on slit is performed automatically with help of a custom written software. In case of faint objects or for special cases, it can be performed also manually with a handpad. Full description of optical elements is provided in Table\,\ref{tblpre}. Furthermore, original Zeiss drawings are available upon request. The overall efficiency of the telescope system is estimated around 0.4 because of the optical elements. The estimated (from optical elements) combined efficiency of OES with the telescope is about 0.05. All useful parameters of the OES system are summarised in Table\,\ref{tblpar}. A more detailed technical description with mechanical setup and all optical elements can be found in the report from the installation phase of the OES \citep{koubsky04}.\newline

\begin{table}
\caption{Parameters for optical elements of OES}             % title of Table
\label{opt}      % is used to refer this table in the text
\centering                          % used for centering table
\begin{tabular}{c c  }        % centered columns (4 columns)
\hline\hline                 % inserts double horizontal lines
Optical element & value   \\    % table heading 
\hline                        % inserts single horizontal line
Beam diameter at collimator (in mm) &  142  \\      % inserting body of the table
Collimator focal ratio  & f/32   \\
Angle of crossdisperser &  $54.5^{\circ}$ \\
Objective focal ratio  & f/1.8   \\
Camera objective lense focal length (mm) & 200   \\
\hline                                   %inserts single line
\end{tabular}
\end{table}

\begin{table}
\caption{Optical elements/parameters of the telescope}             % title of Table
\label{tblpre}      % is used to refer this table in the text
\centering                          % used for centering table
\begin{tabular}{c c c }        % centered columns (4 columns)
\hline\hline                 % inserts double horizontal lines
Element & size   \\    % table heading 
\hline                        % inserts single horizontal line
Telescop focal lenght (m) & 63.5 \\ 
Telescope focal ratio  & f/32 \\ 
M1 mirror shape & parabolic  \\ 
M1 mirror (diam in mm) & 2080  \\ 
M1 focal ratio & f/4.5  \\      % inserting body of the table
M2 mirror (diam in mm)   & 580  \\
M2 shape   & convex hyperbolic  \\
flat ellipt. mirror 1 (mm) &  612$\times$433 \\
flat mirror 2 (diam mm)  & 520   \\
glass plate (mm) & 200   \\
elliptic mirror 3 & $80\times120$  \\
OES slit (mm) & 0.6  \\
\hline                                   %inserts single line
\end{tabular}
\end{table}

\begin{table}
\caption{Instrumental characteristics of OES}             % title of Table
\label{tblpar}      % is used to refer this table in the text
\centering                          % used for centering table
\begin{tabular}{c c c c}        % centered columns (4 columns)
\hline\hline                 % inserts double horizontal lines
Parameter & value &  \\    % table heading 
\hline                        % inserts single horizontal line
%Spectral range & 360--950 nm  \\      % inserting body of the table
slit width (mm) & 0.6 \\
slit width on sky (arcesc) & 2 \\
slit length (arcsec) & 1.8 \\
Echelle (Milton Roy)   & 54.5 g/mm  \\
Blaze angle ($\theta$)   &  $69^\circ$ \\
Spectral resolution   & 51,600 (500 nm)  \\
Linear reciprocal dispersion (\AA/mm) & 2.4 \\
Pixel size $\AA$/pix & 0.0324 \\
pixel size (km/s) & 1.8 \\
spectral range (\AA) & 3753-9195  \\
spectral orders & 56  \\
spectral order number range & 92-36 \\
inter order separation (in pix -blue) & 27\\
inter order separation (in pix - red) & 12 \\
Limiting magnitude ($V$mag) & 13  \\
\hline                                   %inserts single line
\end{tabular}
\end{table}

% \newline
\noindent {\bf Basic data reduction} \newline
 \newline
 Standard spectroscopic calibrations, such as bias frames and lamp flat fields are taken before every observing night. Furthermore, once every seven days a sequence of $10$ dome flat fields is obtained. Cosmic rays are removed based on algorithm by \citet{2001PASP..113.1420V}. We monitor the detector performance once every week by taking the set of ten $30$ minutes dark frames. However, dar subtraction is not required due to low dark current levels. Currently, data reduction is performed with standard scripted IRAF routines\footnote{IRAF is distributed by the National Optical Astronomy Observatories, which are operated by the Association of Universities for Research in Astronomy, Inc., under cooperative agreement with the National Science Foundation.}, after every observing night.   \newline 

\noindent {\bf Influence of wavelength calibrations}\newline

\begin{figure}
   \centering
   \includegraphics[ width=8cm, height=5cm]{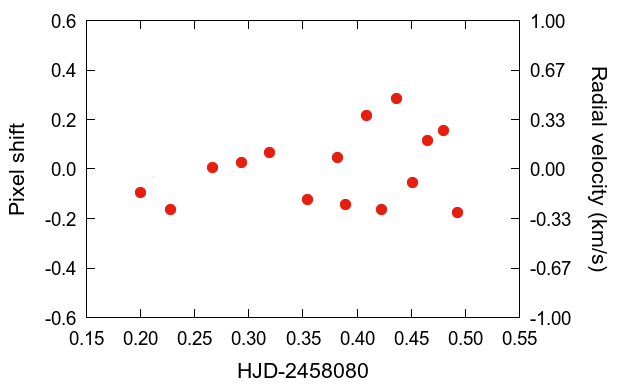}
      \caption{The relative shift of ThAr frames during one night with respect to the reference frame. The scatter is random during the whole night and typically less than 0.5 pixels.
              }
         \label{tharstab}
   \end{figure}

In order to test the stability of the ThAr calibrations (and derive the optimum calibration plan), we performed a test sequence of ThAr frames during a single observing night.  As shown in Fig \ref{tharstab}, the scatter over the course of a night shows no clear trends with random scatter typically less than 0.5 pixels. 

We also investigated potential issues which can influence the stability of the spectrograph. In general, the environment in the Coude room, where the spectrograph is located, is stable down to 1\,deg. We also checked the stability of hollow-cathode lamp calibration frames which were obtained over a period of one year. %Ultimately, we conclude that the ThAr calibration is stable on order of days. 

\begin{figure}
   \centering
   \includegraphics[scale=0.8]{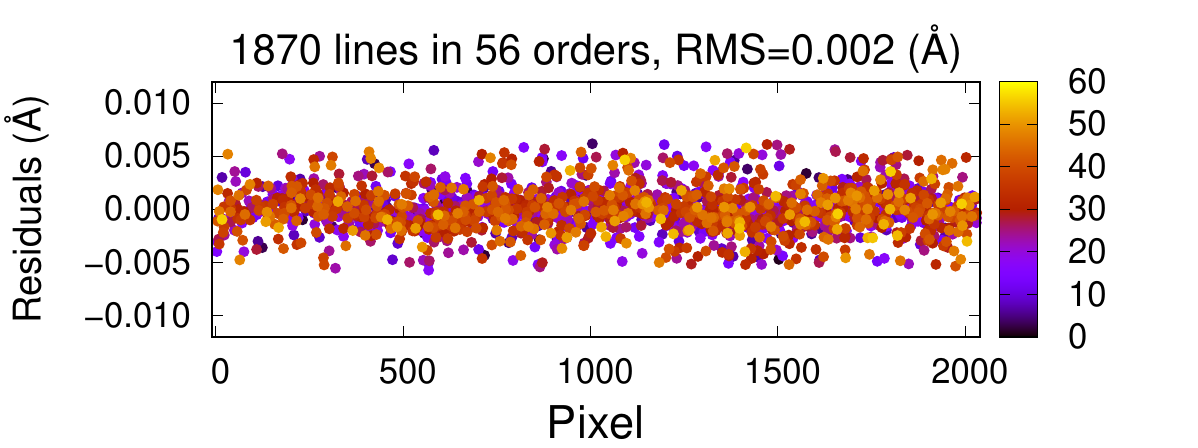}
      \caption{Residuals of the wavelength solution for the OES spectra. Colours represent different orders. The typical residual RMS is 0.002 \AA.
              }
         \label{rms}
   \end{figure}

The wavelength calibration itself is based on the internal atlas from the Coude \'{E}chelle Spectrometer which was installed at 3.6-m telescope at La Silla, Chile. We also cross-checked the calibrations with \citet{redman14, lovis07} and we are able to obtain an RMS of $0.002$\,\AA on the wavelength solution with both ThAr atlases. An example of typical residuals from the wavelength solution is demonstrated in Fig. \ref{rms}. % (radial velocity per dispersion element is $1.67$\,km/s). 

%This RMS corresponds to $3.3$\,m/s (radial velocity per dispersion element is $1.67$\,km/s).

The resulting 1D reduced scientific spectra are stored on local discs and they are shown in local web archive. It is foreseen that, data will be archived at the Virtual Observatory and will be made public after $1$ year of proprietary time, therefore, the whole community can benefit from access to our spectra.

   \begin{figure}
   \centering
   \includegraphics[scale=0.5]{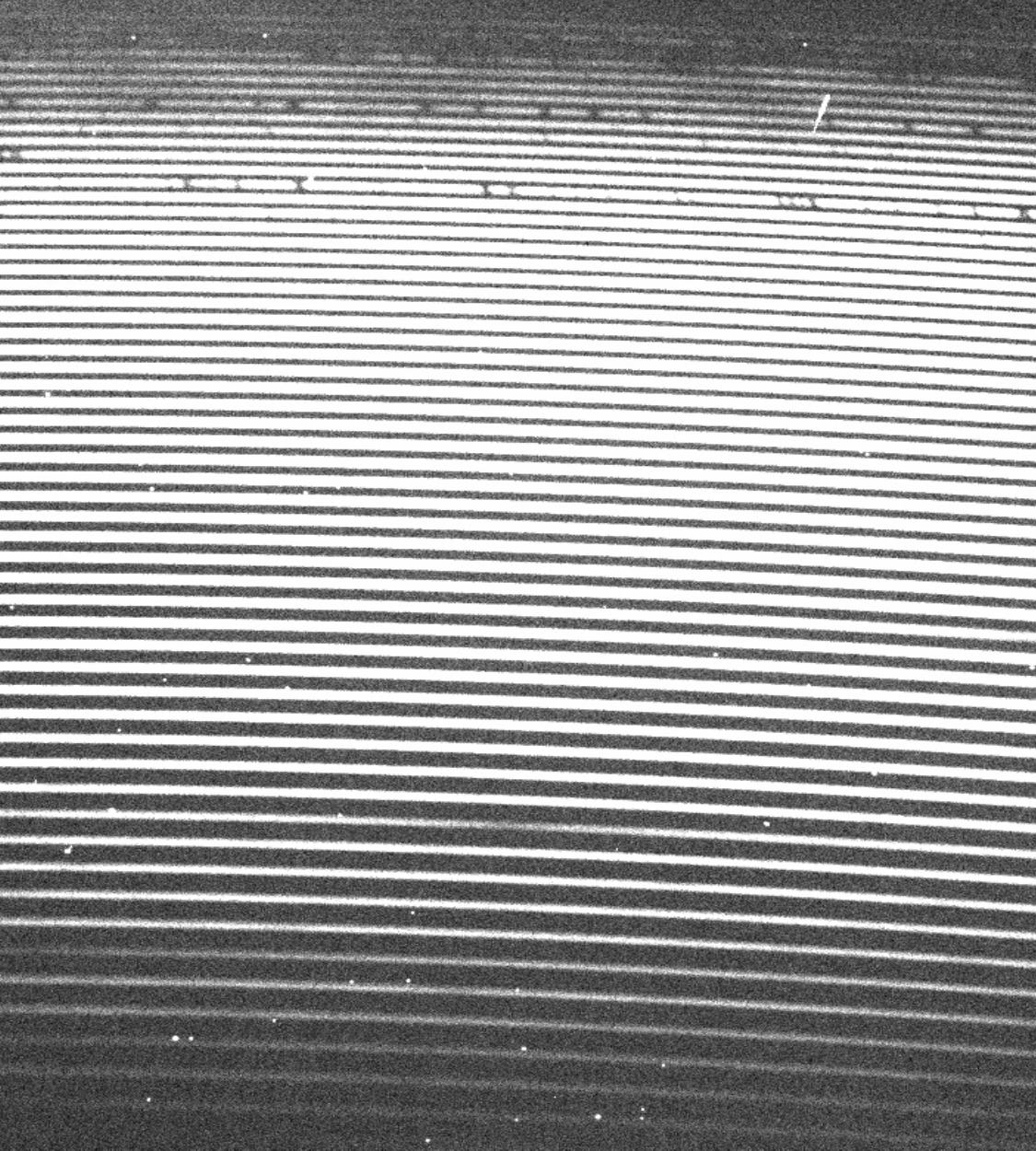}
      \caption{A part of a frame with an \'echelle spectrum of Vega. The reddest orders are at the top.
              }
         \label{starspec}
   \end{figure}

\begin{figure}
   \centering
   \includegraphics[width=8cm, height=5cm]{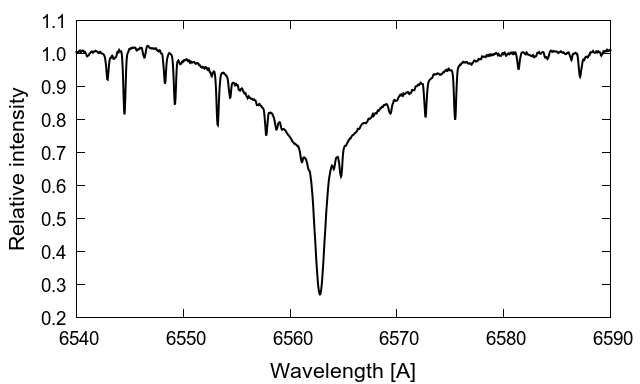}
      \caption{Reduced and extracted spectrum of Vega obtained with OES zoomed around the $H_{\alpha}$ region. 
              }
         \label{reducedspec}
   \end{figure}

For an illustration of the data output, a raw image of a fits frame of Vega is shown in Fig. \ref{starspec}. In the reddest part of the spectra the orders are influenced by some fringing. Reduced stellar 1D spectrum of Vega in the H$_\alpha$ region is shown in Fig. \ref{reducedspec}.

\section{Radial velocity accuracy}\label{sec:RV}

\noindent {\bf HD 109358---stability test of the OES}\newline

\begin{figure}
   \centering
   \includegraphics[scale=0.7]{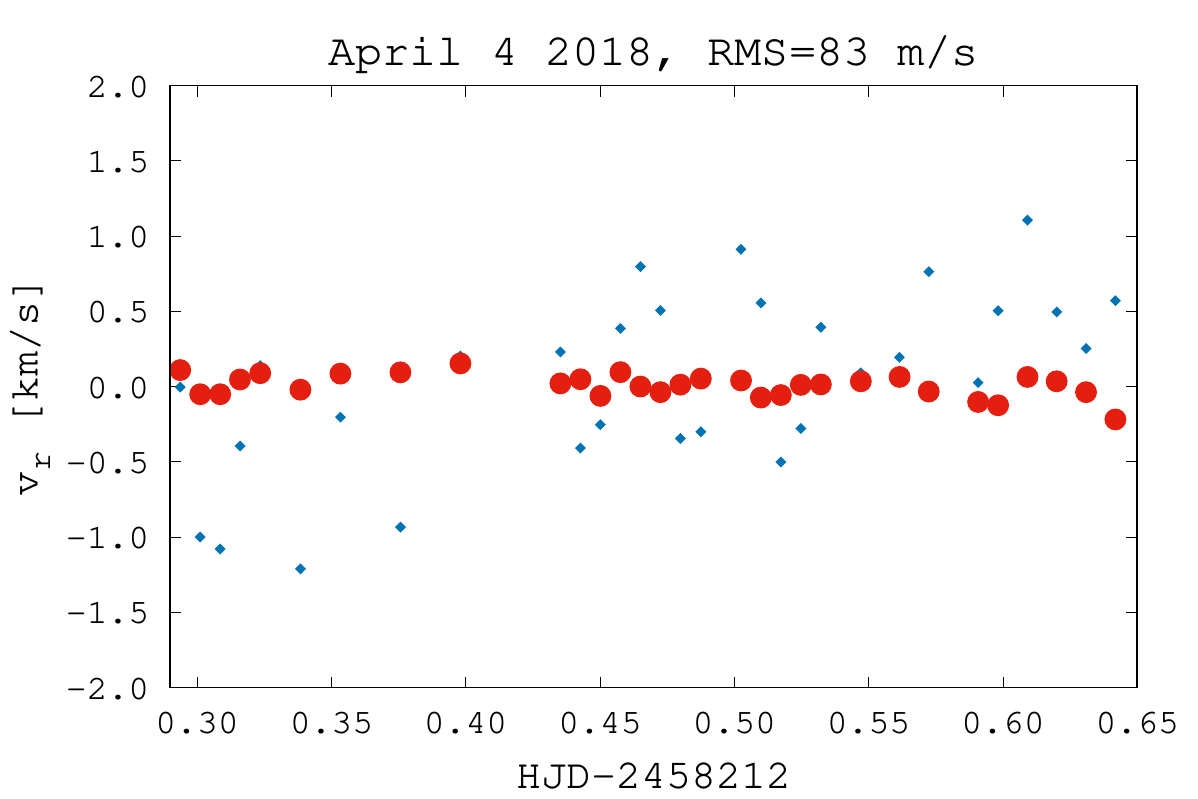}
      \caption{Typical radial velocity curve for one night of spectroscopic time series for standard HD~109358. Data were obtained on one night of 4 April 2018. Red circles represent data where telluric correction was applied, whereas the blue circles represent same data with only ThAr calibration, The RMS of telluric corrected data corresponds to $83$ m/s. RMS of ThAr only calibrated data corresponds to 560 m/s.
              }
         \label{rvcurve}
   \end{figure}

We selected the radial velocity standard star HD~109358 to assess the short term (nightly) and long term (over several days to months) precision that can be obtained with OES.  HD~109358 is a G-type star with $V = 4.25$\,mag included in the Geneva catalogue of radial velocity standards~\citep{udry99}. The star was observed by HARPS and CORALIE and its RV is constant down to a few m/s~\citep{konacki05}. We observed HD~109358 160 times over 15 nights between March 2017 and June 2018, with a typical exposure time of 600 seconds resulting in a signal-to-noise of 50--100 for an extraction aperture of 1 pixel.  The data were reduced following the standard calibration flow using ThAr comparison spectra. To remove the instrumental effects we used the telluric lines present in the wavelength region of $5917--5926$\,\AA. Once the shift of each spectrum with respect to the chosen reference was determined using IRAF routine fxcor, all spectra were shifted for the determined value using IRAF routine dopcor. Figure~\ref{rvcurve} compares however only ThAr calibrated frames (blue circles) and same frames with telluric correction applied (red circles). The scatter in the first part of the RV curve of telluric corrected data in Figure~\ref{rvcurve} is given most likely by the variation of the airmass which was decreasing from 1.5 to 1.02 during observing period.

In the second step, we determined the shifts of stellar lines according to a chosen reference frame, typically a spectrum with highest SNR. These shifts were determined in various wavelength intervals between $4000-5400$\,\AA~which are not affected by telluric lines. The resulting heliocentric radial velocity curve is presented in figure~\ref{rvcurve}. The RMS of radial velocities determined for observations during a single night is roughly 83\,m/s and does not vary between nights. The nightly averages of the measured radial velocities present errors as low as $25$\,m/s. This value corresponds with an integration of approximately $8$ hours and the accuracy is approaching the limit of the telluric line method of RV determination~\citep{gray06}. 

\begin{figure}
   \centering
   \includegraphics[scale=0.68]{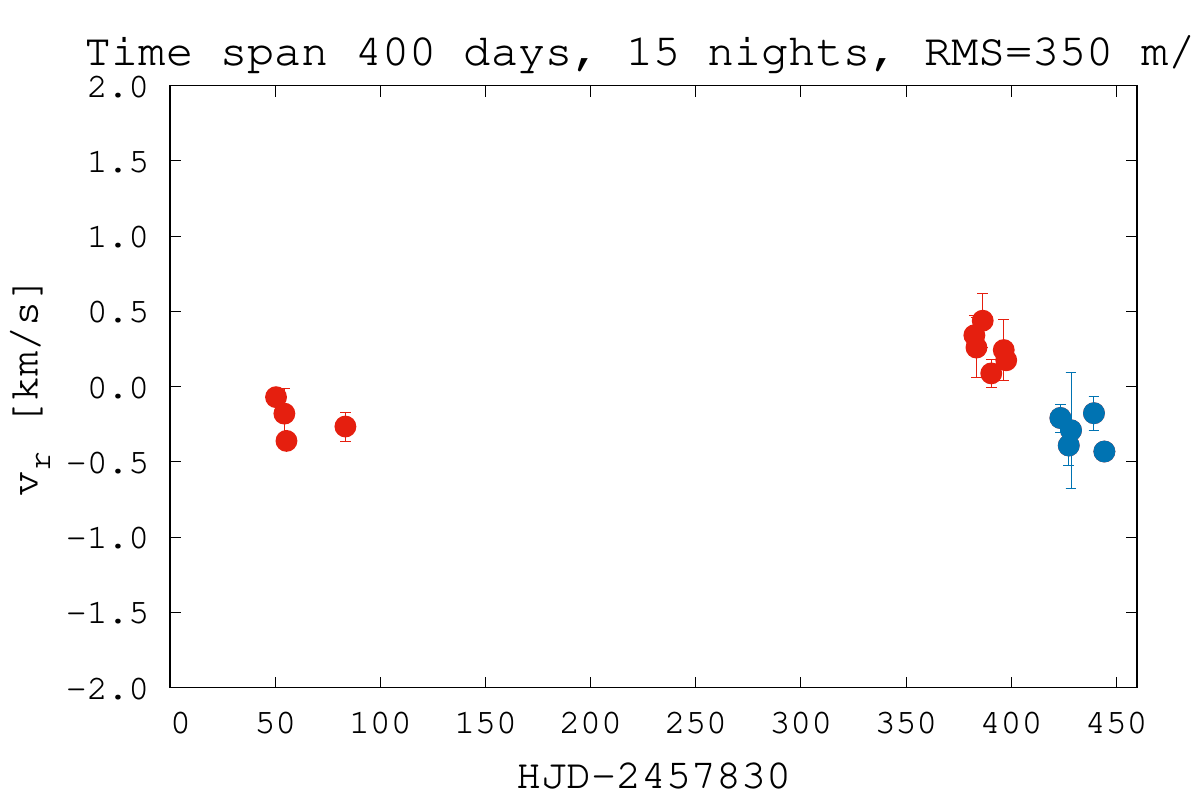}
      \caption{Radial velocity curve for 15 nights of spectroscopic time series for standard HD~109358. Data were taken over a period of 400 days.The RMS corresponds to $350$\,m/s. Blue points indicate measurements of the standard star in 5 nights spanning over three weeks in May to June 2018. The corresponding RMS of blue points only is $111$ m/s.
              }
         \label{rvlong}
   \end{figure}

In the next step, we investigate the long-term stability. The radial velocities were all referenced to a chosen high SNR reference spectrum. The resulting long-term variation is presented in Fig.~\ref{rvlong} with blue points. The measurements span over three weeks and the RMS of the resulting series is 111\,m/s (roughly an order of magnitude lower than the lowest RV precision that would be required for identifying binaries among exoplanetary candidates). Furthermore, measurements of the standard star spanning 450 days have an RMS of $350$\,m/s is presented in the same figure (Fig.~\ref{rvlong}). The reason for outlaying group of measurements around date 370 is not clear and unfortunately there were no data between date 130-350 as the star was not observable. However, the stability of telluric corrected data over one year is still in the order of about $350$\,m/s. RMS of ThAr only calibrated data set is about $1$ km/s. Therefore, we can safely discriminate between planetary candidate and binary or multiple stellar system on short or even longer time scales. Indeed, the precision is sufficient that OES could be used to characterise some hot Jupiter systems which exhibit large RV amplitudes. \newline

% \begin{figure}
%   \centering
 %  \includegraphics[scale=0.38]{g_2018_May-June2018.png}
 %     \caption{Radial velocity curve for 5 nights of spectroscopic time series for standard HD~109358. Data were taken over a period of three weeks in 2018. The RMS corresponds to $111$\,m/s. 
  %            }
  %       \label{rvcurveall}
  % \end{figure}

\noindent {\bf Signal-to-noise and RV accuracies}\newline

\begin{figure}
   \centering
   \includegraphics[scale=0.48]{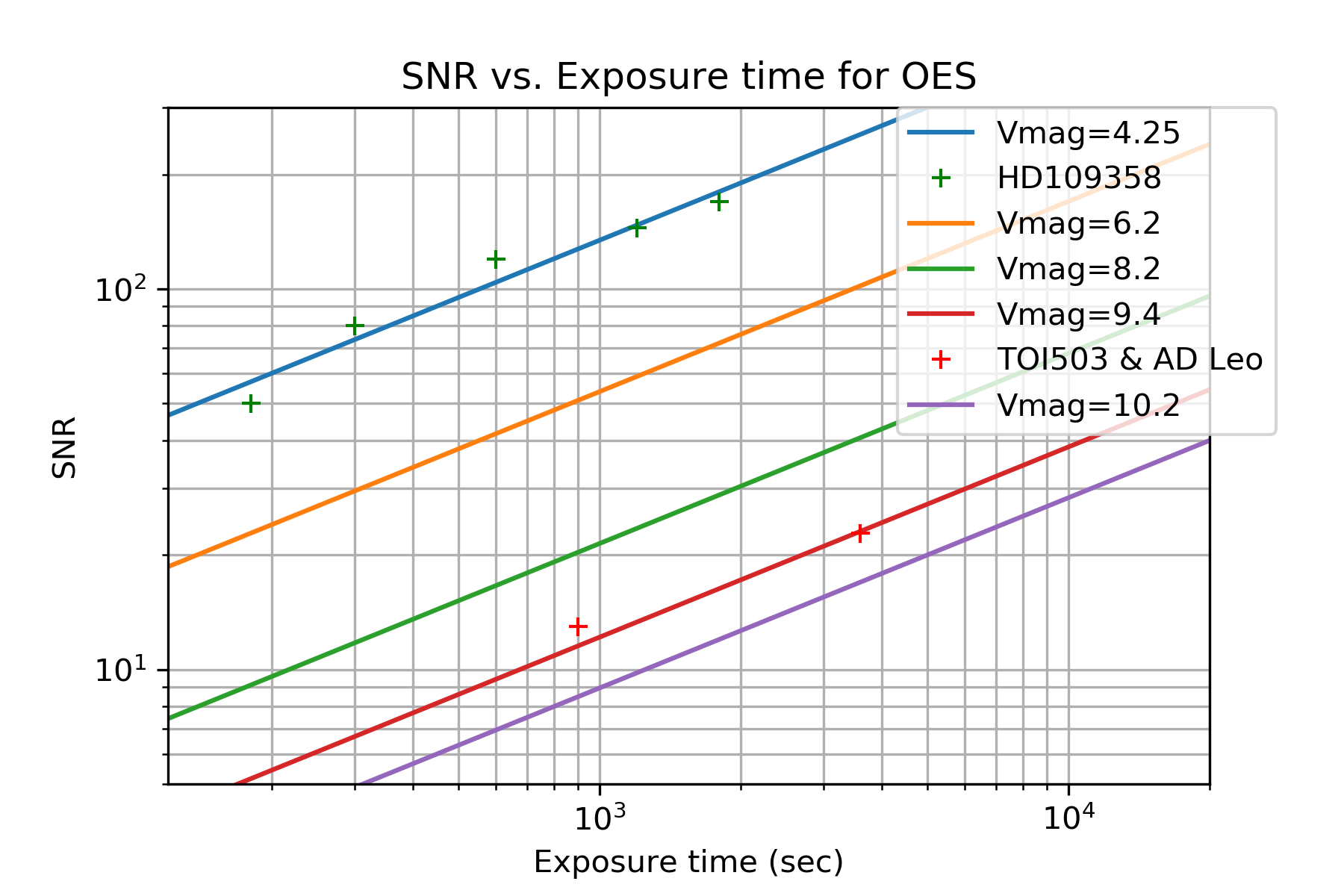}
      \caption{Figure shows theoretical SNR versus exposure time.  Green crosses represent actual measurements of SNR vs exposure time for a star HD109358 (Vmag$=4.25$). Red crosses represent measurements for TOI503 (Vmag$=9.42$) and AD Leo (Vmag$=9.52$) test stars. Colored lines represent different values of Vmag.
              }
         \label{snrgra}
   \end{figure}

Besides the stability of the spectrograph, we tested also the Signal-to-noise (SNR) and exposure time dependence. A graph with SNR versus exposure time for the standard HD109358 is represented with green crosses in Figure\ref{snrgra}. Furthermore, we overplotted also a few more test stars in the same figure. This graph can be used as a guideline for the determination of the necessary exposure. The typical SNR for a star with Vmag=$10$ is SNR $\approx 20$ for an exposure time of 3600 seconds. For a faint test star of Vmag$=12.5$ we obtained a SNR $\approx 7$. 

In addition, we can use our SNR estimates for the determination of the OES RV accuracy. We will apply the values from Figure\ref{rvcurve} to determine instrument characteristic parameters and to deduce the capability of OES. According to \cite{hatzescochran2014} the radial velocity measurement error $\sigma_{RV}$ can be expressed for a (Solar type) star as:
\begin{equation}
\sigma_{RV} = C\times(SNR)^{-1}\times \Delta \lambda^{-0.5}\times R^{-1.5}
\end{equation}
where $SNR$ is the signal-to-noise ratio and $C$ is an instrument specific constant, $\Delta \lambda$ is the wavelength range of the spectrograph and $R$ is the resolving power. For OES, $\Delta \lambda=5900$\,\AA and $R=50,000$. Therefore, we obtain $C=6.5\times10^{12}$. Using the determined value of $C$, we obtain the calculated RV accuracy of about $200-300$ m/s for a Vmag$=9.5$ star. This is consistent with our actual measurements.\newline   

%\newline   

%\newline
\noindent {\bf First results from K2 candidates follow-up}\newline

We present here the first results of K2 candidate follow-up obtained with OES during 2017--2018. We selected 3 suitable K2 candidate stars, EPIC 210925707 (HD 23765), EPIC 206135267 (HD 210809), EPIC 211993818 from K2 campaigns C1 to C6 which were reported in \citet{barros16, pope16, crossfield16}. These targets were chosen based on their brightness (V$<$11) and transit depths (a few percent or more) to ensure that OES was well-equipped to provide accurate and useful RVs.

We derived stellar parameters ($T_{\rm eff}$, metallicity and $\log(g)$ and luminosity class) from the first good spectrum acquired for each of the candidates with iSpec software \citep{cuaresma19} . Subsequently, we obtained a time series of spectra for each star from which an RV curve of the system was determined and orbital parameters obtained (via cross-correlation with a synthetic spectrum with the stellar parameters previously derived). \newline

\noindent {\bf EPIC 210925707}\newline

\begin{figure}
   \centering
   \includegraphics[scale=0.6]{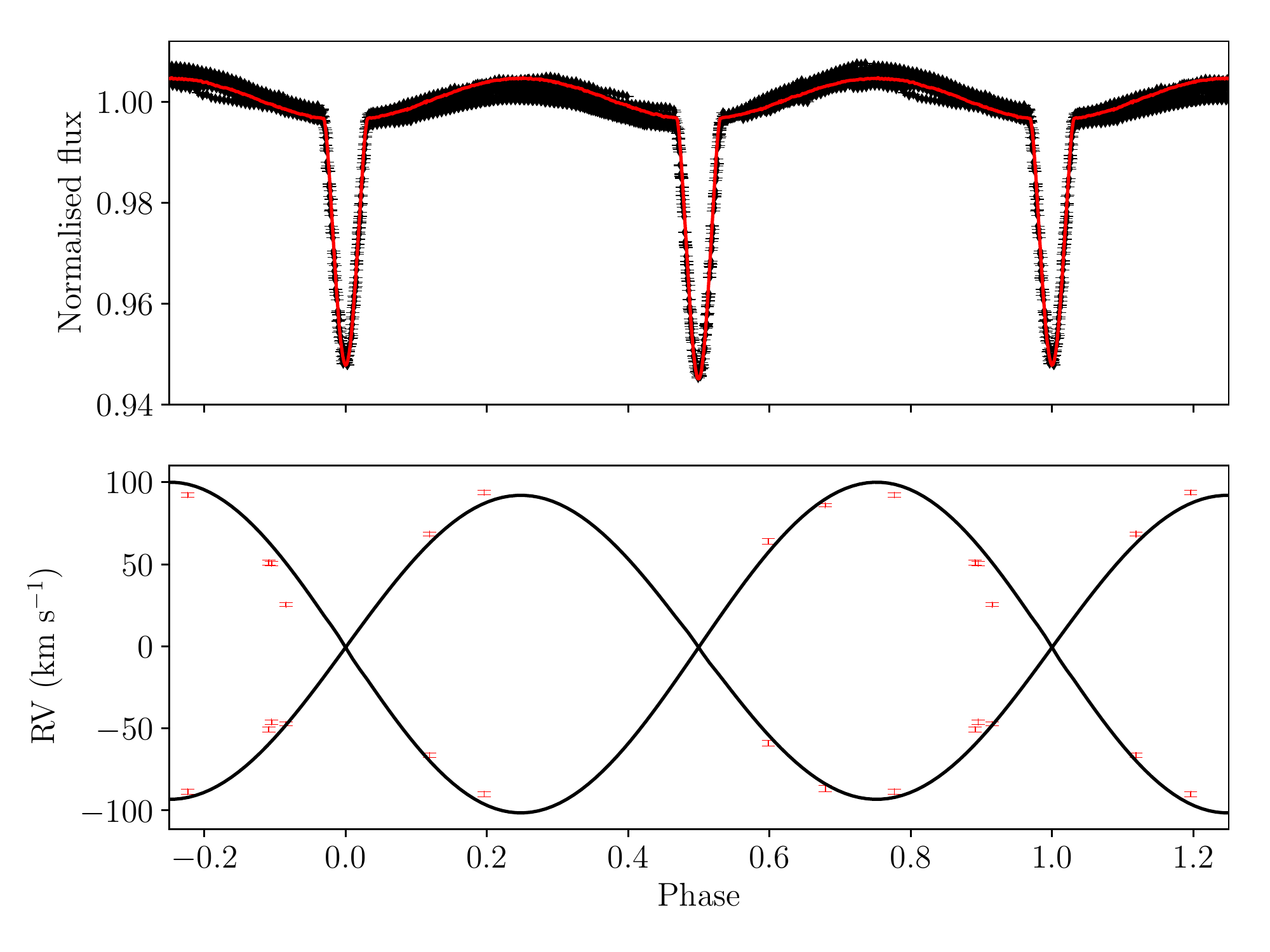}
      \caption{Top panel: Folded K2 light curve of EPIC 210925707 with the \textsc{phoebe 2.0} model overlaid in red. Bottom panel: The folded OES radial velocity measurements along with their model RV curves.
              }
         \label{SB2}
   \end{figure}

EPIC 210925707 (HD 23765) was identified as an exoplanetary candidate from C4 campaign of K2 mission \citep{barros16}. The K2 light curve shows a transit depth $\delta = 4.591 \%$ with a period $1.68658840$ days.  However, the same target was identified as an eclipsing binary in the third version of the Kepler Eclipsing Binary catalog \citep{kirk16}, as the light curve, when phased on double the period identified by \citep{barros16}, seemingly displays two eclipses with slightly different depths and durations.   The K2 light curve is shown phased on a period of 3.3731768 days in the upper panel of Fig. \ref{SB2}.

In total, we obtained 28 spectra of this star across 8 different nights during the 2017-2018 campaign (RVs can be found in Table \ref{tblrv}). The iSpec analysis reveals the presence of two stars of similar spectral type, both roughly F4V (with atmospheric parameters given in table 2).  As such, we were able to derive the radial velocity curve of both stars - with both components showing sinusoidal variability with semi-amplitudes of $97\pm5 $\,km s$^{-1}$ and $101\pm5$\,km s$^{-1}$ (consistent with a mass ratio $\sim$1 and clearly in support of a binary rather than exoplanetary origin for the observed eclipses).  To further constrain the system, the OES radial velocity curve and Kepler light curve were simultaneously modelled using the \textsc{phoebe 2} code \citep{prsa16,horvat18, 2019arXiv191209474J} and following the same technique employed in \citet{boffin18} and \citet{jones19}.  Fitting was performed via a Markov chain Monte Carlo (MCMC) method with the python implementation \textsc{emcee} \citep{foreman-mackey13} and parallelised to run on the LaPalma supercomputer using \textsc{schwimmbad} \citep{price-whelan17}.  The masses, temperature and radii of both stars, as well as the orbital inclination, eccentricity, argument of periastron were all allowed to vary freely around initial estimates based on the spectral type and systemic velocity measured from the OES spectra, and from the shape of the Kepler light curve.  The limb-darkening values for both stars were fixed to the default prescription in \textsc{phoebe 2}, whereby values for each point on the star are interpolated from tables derived from the stellar atmosphere models of \citet{castelli04} as outlined in \citep{prsa16}.

The final \textsc{phoebe 2} model light and radial velocity curves are overlaid on the observations in figure \ref{SB2}, while the derived parameters are listed in table \ref{table:20}.  The model provides a reasonable fit to all observables, although there are residuals between the observed radial velocities and the model consistent with stellar jitter \citep{martinez-arnaiz10}.  Additionally, the observed Kepler light curve shows a clear O'Connell effect (i.e. the out of eclipse brightnesses are asymmetric) which is not replicated by the model.  This is unsurprising as the origin of the effect is uncertain and none of the proposed mechanisms \citep[e.g.\ star spots][]{bell90} are accounted for in the simple binary model.  Ultimately, this will mean that the uncertainties on the measured parameters may be somewhat underestimated.  However, the stellar parameters of the model are found to be in reasonable agreement with those found from the iSpec analysis of the OES spectra, with the temperatures and masses of both stars found to be consistent.  The modelled radii do imply surface gravities greater than one uncertainty away from the value determined by the spectral analysis, however this is one of the most poorly constrained parameters in the iSpec analysis and also, perhaps, the parameter most likely to be affected by the O'Connell effect.  If the effect is caused by the presence of a hot spot or stream on one of the stars \citep{pribulla11}, then the model fitting will tend to over-estimate the radii, therefore underestimating the surface gravities, in order to increase the tidal distortions to match the higher maximum.   Further investigation of the origin of the O'Connell effect in EPIC 210925707 is beyond the scope of this work, but nonetheless the OES observations (and modelling) highlight the aptitude of the instrument for filtering exoplanetary candidates from space missions.\newline

% \begin{figure}
 %  \centering
%   \includegraphics[width=6cm, height=5cm]{HD23765_LC.png}
 %  \includegraphics[width=6cm, height=5cm]{HD23765.png}
  %    \caption{Spectrum of the same star obtained with OES is shown in bottom panel.
 %             }
  %       \label{epic1}
 %  \end{figure}

\begin{table}
\small
\caption{Parameters of EPIC 210925707, EPIC 206135267 and EPIC 211993818}             % title of Table
\label{table:20}      % is used to refer this table in the text
\centering                          % used for centering table
\begin{tabular}{l c c c}        % centered columns (4 columns)
\hline\hline                 % inserts double horizontal lines
Parameter &  EPIC 210925707 & EPIC 206135267 &  EPIC 211993818\\    % table heading 
\hline                        % inserts single horizontal line
RA & $03^{h}48^{m}28^{s}.9308$  &$22^\mathrm{h}13^\mathrm{m}10.7474^\mathrm{s}$  & $08^{h} 24^{m} 49^{s}.1841$\\      % inserting body of the table
DEC   &  $+21^{\circ}47^{\prime}50.8233^{\prime\prime}$  & $-11^{\circ}10^{\prime}38.4888^{\prime\prime}$& $+20^{\circ} 09^{\prime} 10.7633^{\prime\prime}$\\
$V_{\rm Kepler}$   & 9.52  & 9.23\&  7.28\\
\multicolumn{4}{c}{\bf iSpec parameters}\\
$T_{\rm eff}$ (\,K)   & $6300 \pm 112$ & $5071 \pm 163$ & $5393 \pm 108$\\
log(g) &  $4.64 \pm 0.38$ & $3.96 \pm 0.39$& $2.58 \pm 0.3$\\
Fe/H &  $-1.4 \pm 0.1$ & $-0.86 \pm 0.13$ & $-0.45 \pm 0.3$  \\
Spectral type &  F5V & G8IV  & G5III \\
\multicolumn{4}{c}{\bf \textsc{phoebe 2} parameters}\\
Period (days) &  3.3731768   & $2.57340205$ (fixed) & $3902\pm 3$ (fixed)\\
Time of conjunction (HJD) & 2,457,062.23826 (fixed) & $2,456,982.6207794$ (fixed) & $2451306 \pm 8$ (fixed) \\
Orbital inclination ($^\circ$) & $76.20\pm0.04$ & N/A & N/A\\
Eccentricity & $0.0057 \pm 0.0004$  & $0.22 \pm 0.07$ & $0.612 \pm 0.03$ \\
Semi-amplitude(s) (km/s) & $ 97\pm 5 $, $101\pm 5 $ & $13.3 \pm 1.9$ & $18.7 \pm 0.1 $\\
%Systemic velocity (km/s ) & $-0.7 ^{+0.6}_{-0.7}$ & $-2.1 \pm 1.4$\,km/s & $-1.07 \pm 0.05$ \\
RMS of O-C (km/s) &9.06& 2.88 N/A \\
Argument of periastron (radian) & $1.70\pm0.03$& $3.05\pm 2.9$&   $5.42 \pm 0.1$  \\
$T_{\rm 1 eff}$ (K) & $6480^{+20}_{-30}$ &&\\
$T_{\rm 2 eff}$ (K) & $6710\pm20$ &&\\
$R_{\rm 1}$ (R$_\odot$) & $1.95^{+0.03}_{-0.02}$ & & \\
$R_{\rm 2}$ (R$_\odot$)& $2.08^{+0.03}_{-0.04}$&&\\
$M_{\rm 1}$ (M$_\odot$) & $1.33^{+0.02}_{-0.04}$ &&\\
$M_{\rm 2}$ (\,M$_\odot$)& $1.44\pm0.06$&&\\
\hline                                   %inserts single line
\end{tabular}
\end{table}

\begin{table}
\caption{Radial velocities from OES for EPIC 210925707}             % title of Table
\label{tblrv}      % is used to refer this table in the text
\centering                          % used for centering table
\begin{tabular}{c c c }        % centered columns (4 columns)
\hline\hline                 % inserts double horizontal lines

& star 1&   star 2  \\   % table heading 
Julian Date (day) &   RV+error (km/s)  &   RV+error (km/s) \\  

\hline                        % inserts single horizontal line

2457983.51612     &  $ -66.4 \pm1.5$    &   $ 68.5 \pm1.3$  \\ 
2457988.50698     &   $63.9 \pm1.7$    &    $-59.1 \pm1.9$ \\ 
2457989.50869    &    $50.5 \pm1.2$    &    $-46.2 \pm1.3 $\\ 
2457990.52336     &   $-90.1 \pm1.5$    &    $93.8 \pm1.4$ \\ 
2457992.48173    &   $92.2 \pm1.4$    &    $-88.6 \pm1.5$ \\ 
2458005.64496    &    $86.0 \pm1.0$     &   $-86.6 \pm2.0 $\\ 
2458016.47976    &    $51.0 \pm1.5$    &    $-50.6 \pm1.6 $\\ 
2458043.54726     &   $25.4 \pm1.2$    &    $-47.4 \pm1.2 $\\ 

\hline                                   %inserts single line
\end{tabular}
\end{table}

\noindent {\bf EPIC 206135267}\newline

\begin{figure}
   \centering
   \includegraphics[scale=0.5]{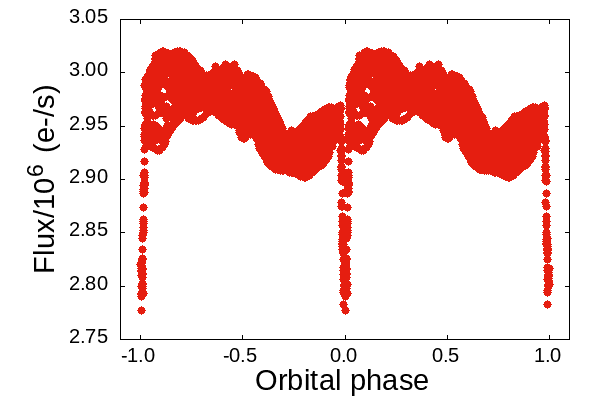}
      \caption{K2 light curve of EPIC 206135267.
              }
         \label{spect2}
   \end{figure}

The star EPIC 206135267 (HD 210803) was identified as an exoplanetary candidate in C3 showing transit depth $\delta = 4.986\,\%$ and a period $P=2.57340205$ days by \citet{barros16}. This system was indeed resolved as a binary from observations from KECK AO with NIRC2 \citet{schmitt16}. The original K2 light curve is shown in Fig.\,\ref{spect2}.

\begin{figure}
   \centering
      \includegraphics[scale=0.5]{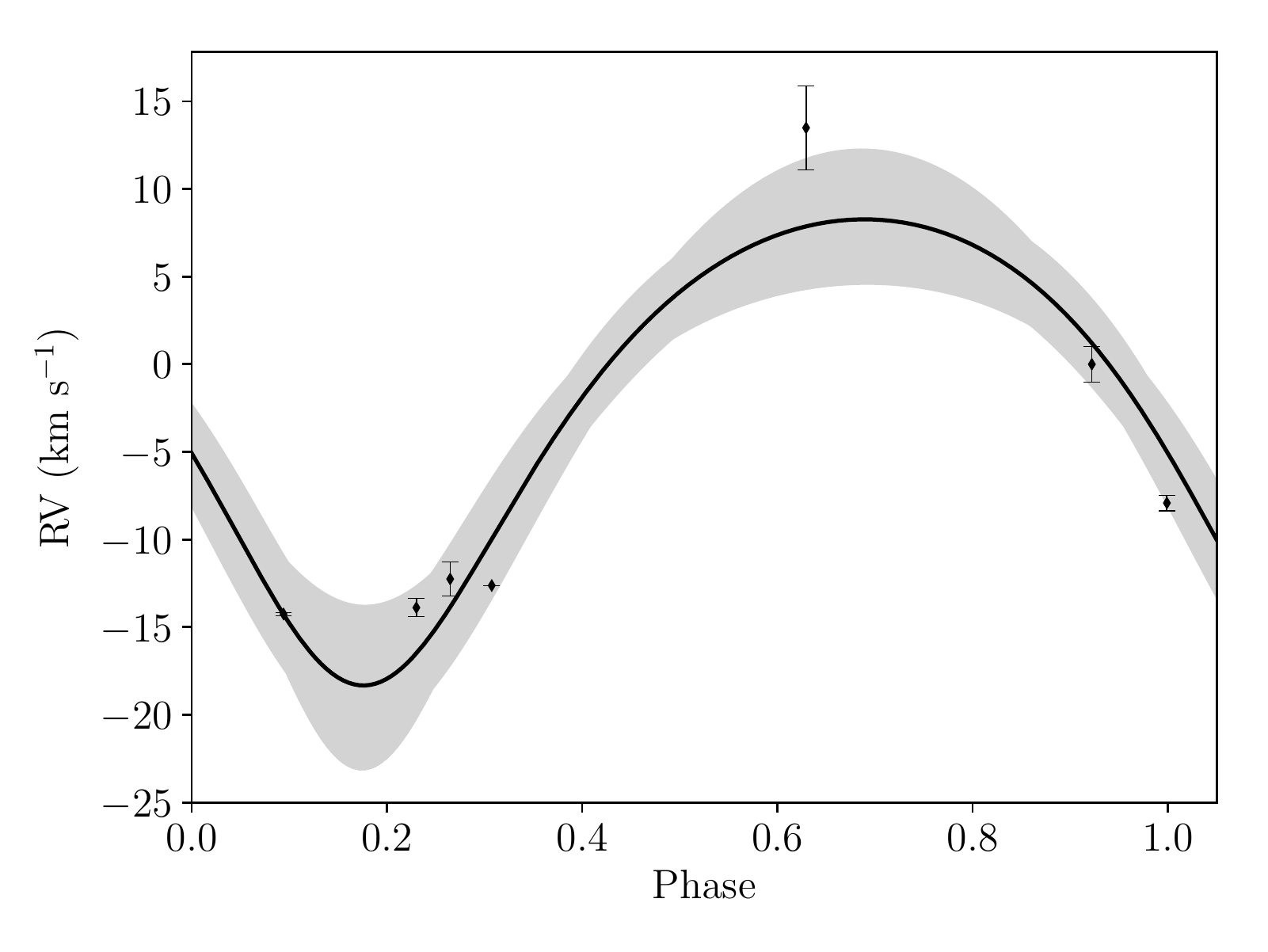}
      \caption{OES RV measurements for EPIC 206135267 with the fit and $1\sigma$ uncertanties indicated by the grey band.
              }
         \label{spec12}
   \end{figure}

For this star, we obtained 7 usable spectra across 7 different nights during the 2017--2018 (RVs are presented in Table \ref{tblrv2}). We determined the stellar spectral type as G8IV star. The RV curve of this star obtained with OES (see Fig.\,\ref{spec12}) presents variations with a semi-amplitude $K=13.3 \pm 1.9$ km/s, where period for fitting was fixed to value of $P=2.57340205$ days which was determined by K2 team \citep{barros16}. The variations in our radial velocities and spectral type clearly identify this star as a binary. Our measurements thus confirm the results from high-resolution AO observations and derive orbital parameters for this system (summarized in Table \ref{table:20}). \newline

\begin{table}
\caption{Radial velocities from OES for EPIC 206135267}             % title of Table
\label{tblrv2}      % is used to refer this table in the text
\centering                          % used for centering table
\begin{tabular}{c c  }        % centered columns (4 columns)
\hline\hline                 % inserts double horizontal lines

Julian Date (day) &   RV+error (km/s) \\  

\hline                        % inserts single horizontal line

2457953.47468   &    $ -12.3\pm1.0$    \\ 
2457983.47368   &    $ 0.0 \pm1.0 $       \\ 
2457989.41321   &    $ -13.9 \pm0.5$  \\ 
2457990.44055   &    $ 13.5 \pm2.4$\\ 
2457991.3917    &    $ -7.9 \pm0.5$\\ 
2458015.34591   &    $ -12.6 \pm0.0$\\ 
2458017.37047  &     $ -14.3 \pm0.1$\\ 

\hline                                   %inserts single line
\end{tabular}
\end{table}

%\begin{table}
%\caption{Parameters of EPIC 206135267}             % title of Table
%\label{table:5}      % is used to refer this table in the text
%\centering                          % used for centering table
%\begin{tabular}{c c }        % centered columns (4 columns)
%\hline\hline                 % inserts double horizontal lines
%Parameter & value   \\    % table heading 
%\hline                        % inserts single horizontal line
%RA & $22^\mathrm{h}13^\mathrm{m}10.7474^\mathrm{s}$ \\      % inserting body of the table
%DEC   &  $-11^{\circ}10^{\prime}38.4888^{\prime\prime}$\\
%$V_{\rm Kepler}$   &  9.23\,mag  \\
%\multicolumn{2}{c}{iSpec parameters}\\
%$T_{\rm eff}$    &   $5071 \pm 163$\,K  \\
%log(g) & $3.96 \pm 0.39$ \\
%Fe/H &  $-0.86 \pm 0.13$ \\
%Spectral type &  G8IV \\
%\multicolumn{2}{c}{RV fit parameters}\\
%Period &  $2.57340205$\,d (fixed) \\
%Time of conjunction & $2,456,982.6207794$\,JD (fixed) \\
%Eccentricity & $0.22 \pm 0.07$   \\
%Semi-amplitude & $13.3 \pm 1.9$\,km/s \\
%Systemic velocity (gamma) & $-2.1 \pm 1.4$\,km/s  \\
%Argument of periastron & $3.05\pm 2.9$\,radian  \\
%\hline                                   %inserts single line
%\end{tabular}
%\end{table}

\noindent {\bf EPIC 211993818}\newline

\begin{figure}
   \centering
   \includegraphics[scale=0.5]{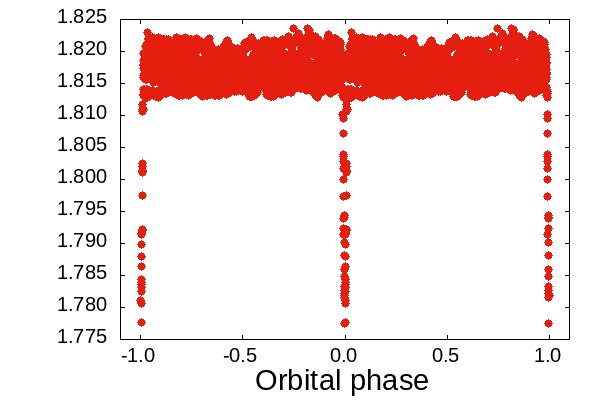}
      \caption{Original K2 light curve of EPIC 211993818. %(top) and the spectrum of the star used for determination of the spectral type (bottom).
              }
         \label{spect3}
   \end{figure}

The main parameters for EPIC 211993818 (HD 70826) from the C5 campaign of K2 mission are summarised in Table \ref{table:20}. We selected this system because it was identified as an exoplanetary candidate at the time of our selection \citet{pope16}, showing a transit depth $\delta = 1.9\,\%$ and a period $P_{\rm component} = 8.9870246$ days and later still kept as a candidate by \citet{mayo18}. Fig. \ref{spect3} shows the K2 light curve. We obtained 9 usable spectra with OES for this star across 9 different nights, from which we identified the luminosity class of the star as G5III giant. Our RV curve of this star presents no apparent RV variations down to 630 m/s (our RV measurements are presented in Table \ref{tblrv3}). 

\begin{figure}
   \centering
     \includegraphics[width=9cm, height=5cm]{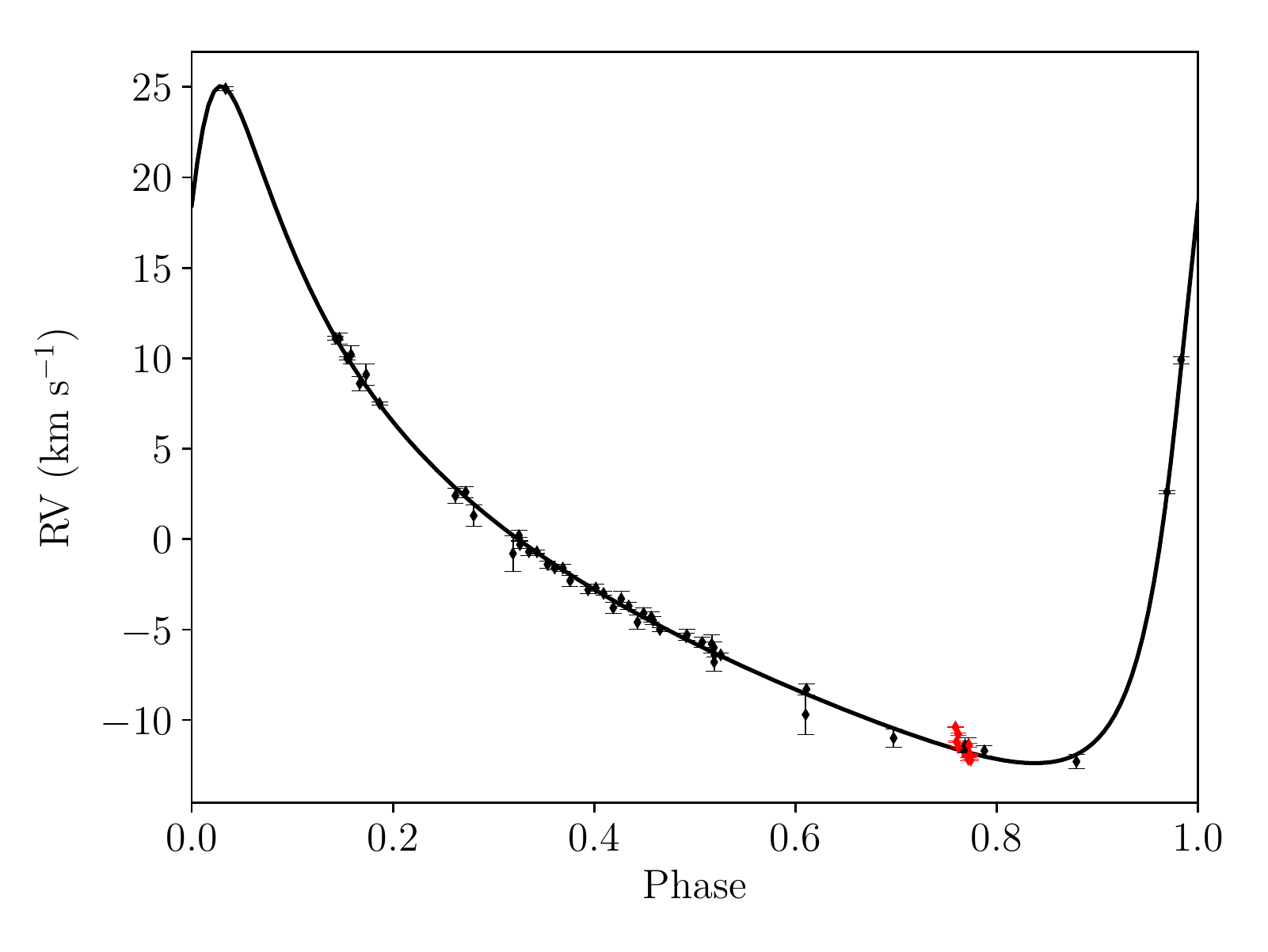}
      \caption{The \citet{carquillat05} RV curve of EPIC 211993818 with new OES data marked as red points showing the good agreement between the measurements. 
              }
         \label{spect4}
   \end{figure}

However, in the literature, we found that this system was identified as multiple system of SB1 type with a second stellar component most likely of A spectral type having a period of $P=3,902$ days. The OES data compared to those from other projects \citep{carquillat05} are shown in Fig. \ref{spect4}. The exact configuration of the system at time of discovery was unclear. The new OES measurements are consistent with previous points obtained by  \citet{carquillat05} as is the subsequent fit to the combined dataset. 

Data from K2 confirm that the system is a three body system with the detected transit occurring on the smaller component of the triple. We estimated the radius $R \approx 11.5$\,\(\textup{R}_\odot\) and luminosity $L \approx 101$\,\(\textup{L}_\odot\) of the primary based on the Gaia data archive, implying a mass of around $M \approx 3$\,\(\textup{M}_\odot\). If this information is combined with K2 light curve data and we assume that the most of the light comes from the primary G5III star, then the real (uncontaminated) transit depth would be too large for the system to be a planet. This is also consistent with conclusions of \citep{carquillat05}, however, now we can confirm that this object is really a triple star and we can also clearly reject the planetary candidate scenario.  We assume that the reason for \citet{pope16} and \citet{mayo18} to include the star in the candidate list is the fact that they were initially not aware of existing spectroscopic data sets. Large number of candidates can not be immediately cross-correlated with those existing data sets.

\begin{table}
\caption{Radial velocities from OES for EPIC211993818}             % title of Table
\label{tblrv3}      % is used to refer this table in the text
\centering                          % used for centering table
\begin{tabular}{c c  }        % centered columns (4 columns)
\hline\hline                 % inserts double horizontal lines

Julian Date (day) &   RV+error (km/s) \\  

\hline                        % inserts single horizontal line

2458170.47868   &    $-10.39\pm0.01$\\
2458174.43933    &    $-11.20\pm0.06$\\
2458181.44328   &     $-10.78\pm0.05$\\
2458182.449    &    $-11.49\pm0.05$\\
2458220.29247 &      $ -12.17\pm0.10$\\
2458223.31822   &    $ -11.39\pm0.10$\\
2458227.35398   &     $-11.90\pm0.03$\\
2458226.35478   &     $-11.95\pm0.06$\\
2458229.363535 &      $ -12.21\pm0.05$\\

\hline                                   %inserts single line
\end{tabular}
\end{table}

%\begin{table}
%\caption{Parameters of EPIC 211993818}             % title of Table
%\label{tabepic}      % is used to refer this table in the text
%\centering                          % used for centering table
%\begin{tabular}{c c }        % centered columns (4 columns)
%\hline\hline                 % inserts double horizontal lines
%Parameter & value   \\    % table heading 
%\hline                        % inserts single horizontal line
%RA &  $08^{h} 24^{m} 49^{s}.1841$ \\      % inserting body of the table
%DEC   & $+20^{\circ} 09^{\prime} 10.7633^{\prime\prime}$ \\
%$V_{\rm Kepler}$   &  7.28\,mag \\
%\multicolumn{2}{c}{iSpec parameters}\\
%$T_{\rm eff}$    &  $5393 \pm 108$\,K \\
%log(g) &  $2.58 \pm 0.3$ \\
%Fe/H &  $-0.45 \pm 0.3$ \\
%Spectral type &  G5III \\
%\multicolumn{2}{c}{RV fit parameters}\\
%Period &  $3902\pm 3$\,d (fixed) \\
%Time of conjunction & $2451306 \pm 8$\,JD (fixed) \\
%Eccentricity & $0.612 \pm 0.03$   \\
%Semi-amplitude & $18.7 \pm 0.1 $\,km/s \\
%Systemic velocity (gamma) & $-1.07 \pm 0.05$\,km/s  \\
%aArgument of periastron & $5.42 \pm 0.1$\,radian  \\
%\hline                                   %inserts single line
%\end{tabular}
%\end{table}

\section{Conclusions}\label{sec:conc}

We have presented the OES instrument and first results from our science verification run between 2017 and 2018.  Our key program for OES comprises the ground-based follow-up of exoplanetary candidates from space missions, especially TESS and later PLATO. We have demonstrated that the OES spectrograph is stable on the order of a few weeks down to 80--100 m/s and over the whole year to about 300 m/s. Therefore, we are able to use the instrument for the rejection of false positives and thus filtering of exoplanetary candidates for more-detailed follow-up with larger facilities -- our primary objective. We can securely rule out binary and multiple stellar systems, brown dwarf systems and flag potential background binaries. 

Weather at our location restricts the useable observing time to about 30\,$\%$ of total telescope time, corresponding to slightly more than 100 nights available for our group. The spectroscopic follow-up of space missions on 2-m telescopes will be crucial for the mission success. If we take into account the weather conditions and the handful of other existing 2-m class telescopes across Europe, then the probability of a clear night in at least one location will be dramatically increased and thus such a network would have a similar yield as a new telescope installed at an observing site with excellent conditions. Furthermore, as most are equipped with similarly capable instrumentation, there is no need to build new instruments for these telescopes.

Since 2017, as a part of the science verification program of OES, we are pursuing a joint Ond\v{r}ejov-Tautenburg RV follow-up program of exoplanetary candidates from {\it Kepler}/K2 missions {as reported in \citet{2019MNRAS.tmp.2155S}, where we introduced first joint analysis of data obtained during simultaneous monitoring of A stars with brightness range of Vmag$=7-11$ from {\it Kepler} which were reported as systems with planetary candidates in \citet{balona}. Typical exposure times were dependent on brightness but covered the range of 1800-3600 seconds. However, we could not confirm any planet but we were able to draw conclusions about the lack of planets around A stars. This particular work also highlights the strength of two 2-m telescopes working together. Another such example of joint observing campaign was presented in \citet{2019MNRAS.484.4352G} where a system Kepler-410 (Vmag$=9.5$) was studied together with spectrographs at Tatransk\'{a} Lomnica, Slovakia. Typical exposure times with OES were in this case 45 minutes. 

We are currently monitoring TESS candidates. In addition to the RV follow-up, we will be able to provide initial stellar parameters for monitored stars by the space missions. We will be thus able to contribute also to e.g.; the PLATO input catalogue - a particularly important task as transit photometry allows only to determine the ratio of the star to planet radii.  This means that we have to know the radii of the stars in order to determine the radii of the planets. Our goal is to obtain the R$_*$, M$_*$ from $T_{\rm eff}$, $\log(g)$, and the abundances, determined from typically 2--3 high resolution spectra \citep{guenther09}.  Another interesting science case for 2-m telescopes was presented in our earlier paper \citep{2019PASP..131h5001K}, where we discuss the use of 2-m telescopes for characterisation of TESS targets. In the flood of data and newly-confirmed exoplanets around bright stars, 2-m class telescopes could potentially even characterise exoatmospheres of selected targets.

In the course of the latest stage of science verification process, we were able to collect spectroscopic data for 3 selected candidates from K2 space mission. Our measurements confirmed the nature of the candidates as binaries and provided new orbital elements for all of them. Our first results show that we will also be able to characterise some close-in and hot Jupiter-mass planetary systems from e.g. TESS which will exhibit large semi-amplitudes of few hundreds of m/s. As service for community, the OES data are currently publicly available upon a request. The first scientific results from OES were reported in \citet{2019MNRAS.487.4230S} presenting a discovery of a first Ap star (Vmag$=8.16$) with large stellar spots and exhibiting $\delta$ Scuti pulsations. Furthermore, the primary Ap star HD99458 is orbited by the red dwarf companion. This system was reported originally as a planetary candidate from K2 mission. We took a series of several dozens of measurements mostly with 1800 seconds exposure times. In our recent paper \citet{Subjak}  based on OES data, we report on the confirmation of the first transiting Brown dwarf detected with TESS space mission. Our team led joint analysis of data obtained with OES and with other mid-sized aperture telescope facilities and determined the mass and radius of the Brown dwarf TOI-503b. The apparent brightness of TOI-503 is Vmag$=9.40$ and typically the exposures with OES were about 2700 seconds long.

 It is worth to compare OES performance with similar class instrumentation on similar aperture telescopes, therefore, we chose a representative sample consisting of SOPHIE, HERMES and Tautenburg spectrographs. The spectrograph SOPHIE at 1.93-m telescope at Observatoire de Haute Provence has two modes with $R_R=40000$ and with $R_E=75000$ with wavelength range coverage of 387 to 694 nm. The difference to OES is that SOPHIE is a fibre fed spectrograph located in the thermally controlled chamber. The performance of SOPHIE for RV accuracies is down to 50 cm/s short term and about 3-4 m/s in a month \cite{2011SPIE.8151E..15P, 2013A&A...549A..49B}. 

HERMES at 1.2-m Mercator telescope at La Palma, Spain, is a fiber fed spectrograph with spectral resolving power of $R=85000$. HERMES is located in thermally well controlled room, therefore the RV accuracies are around 2 m/s \citep{raskin}. The Tautenburg \'{E}chelle Spectrograph (TLS) is mounted at 2-m Alfred Jensch telescope in Thuringia, Germany. The spectral resolving power R$=67,000$ and the reported best accuracy in RVs is about 1.7 m/s for a very bright star $\beta$ Gem \citep{2006A&A...457..335H}. However, the measurements were taken with an Iodine cell. OES is not thermally stabilised as SOPHIE and HERMES and it is not a fiber fed spectrograph, therefore, it is only directly comparable with TLS. The accuracy of OES without Iodine cell is a factor $1.2-1.5$ worse than TLS (as seen in e.g. \citet{Subjak}, figure 5). This is given by the optical design of OES with not optimal camera objective. However, OES was planned as a spectrograph for stellar physics of hot Be stars and not for precise radial velocity measurements. The strengths of OES are joint observing with TLS and the availability of the telescope time. In light of our science verification results and due to need to use the OES for exoplanetary science, upgrades of the OES, especially thermal stability and the optical design (objective exchange) can be planned in future to improve OES performance.

Finally, we plan to install a new \'{e}chelle spectrograph, PLATOSpec, at ESO La Silla 1.52-m telescope, which will be fully dedicated to TESS and PLATO ground-based follow-up observations and which should become operational in 2022. Combined facilities at Ond\v{r}ejov, Tautenburg and La Silla will be a powerful team to perform a ground-based support observations for future exoplanetary space missions from both hemispheres. \newline

\ack 

\noindent We would like to thank to an anonymous referee for suggestions which significantly improved this paper. PK, JS and MB would like to acknowledge the support from GACR international grant 17-01752J and some travel cost to visit collaborators from ERASMUS+ grant Agreement no. 2017-1-CZ01-KA203-035562. EG and SS acknowledge the DFG grant GU 646/20. DJ acknowledges support from the State Research Agency (AEI) of the Spanish Ministry of Science, Innovation and Universities (MCIU) and the European Regional Development Fund (FEDER) under grant AYA2017-83383-P.  DJ also acknowledges support under grant P/308614 financed by funds transferred from the Spanish Ministry of Science, Innovation and Universities, charged to the General State Budgets and with funds transferred from the General Budgets of the Autonomous Community of the Canary Islands by the Ministry of Economy, Industry, Trade and Knowledge. MS acknowledges the OP VVV project Postdoc@MUNI.cz No. CZ$.02.2.69/0.0/0.0/16\_027/0008360$. We acknowledge the use of IRAF and SIMBAD. We would like to thank to Dr. David Latham for his comments which greatly improved this article. This work has made use of data from the European Space Agency (ESA) mission
{\it Gaia} (\url{https://www.cosmos.esa.int/gaia}), processed by the {\it Gaia}
Data Processing and Analysis Consortium (DPAC,
\url{https://www.cosmos.esa.int/web/gaia/dpac/consortium}). Funding for the DPAC
has been provided by national institutions, in particular the institutions
participating in the {\it Gaia} Multilateral Agreement.  The authors thankfully acknowledge the technical expertise and assistance provided by the Spanish Supercomputing Network (Red Espa\~nola de Supercomputaci\'on), as well as the computer resources used: the LaPalma Supercomputer, located at the Instituto de Astrofs\'ica de Canarias.

\newcommand{\newblock}{}

\end{document}